
\magnification=1200
\baselineskip=22pt
\centerline{\bf Do Experiments and Astrophysical
Considerations
Suggest an}
\centerline{\bf  Inverted Neutrino Mass Hierarchy?}
\centerline{\sl George M. Fuller}
\centerline{Department of Physics, University of California,
San Diego,
La Jolla, CA 92093-0319}
\centerline{\sl Joel R. Primack}
\centerline{Department of Physics,
University of California, Santa Cruz, Santa Cruz, CA 95064}
\centerline{and}
\centerline{\sl Yong-Zhong Qian}
\centerline{Institute for Nuclear Theory, NK-12, University
of
Washington, Seattle, WA 98195}
\baselineskip=24pt plus 2pt
\centerline{\sl ABSTRACT}

The recent results from the Los Alamos neutrino oscillation
experiment, together with assumptions of neutrino
oscillation
solutions for the solar and atmospheric neutrino deficit
problems,
may place powerful constraints on any putative scheme for
neutrino
masses and mixings.
Assuming
the validity of these experiments and
assumptions, we argue that a nearly unique spectrum of
neutrino masses
emerges as a fit, if two additional astrophysical arguments
are
adopted:  (1) the sum of the light neutrino masses is  $\sim
5\ {\rm
eV}$, as large scale structure
simulations with mixed cold plus hot dark matter seem to
suggest; and
(2) $r$-process nucleosynthesis
originates in neutrino-heated ejecta from Type II
supernovae. In this
fit, the masses of the neutrinos must satisfy $m_{{\nu}_e}
\approx
m_{{\nu}_s} \approx 2.7\ {\rm eV}$ (where ${\nu}_e$ is split
from a
sterile species, ${\nu}_s$, by $\sim {10}^{-6} \ {\rm eV}$)
and
$m_{{\nu}_{\tau}} \approx m_{{\nu}_{\mu}} \approx 1.1\ {\rm
eV}$
(where these species are split by $\sim {10}^{-2} \ {\rm
eV}$). We
discuss
alternative neutrino mass spectra that are allowed
if we decline to adopt certain experiments or astrophysical
models.

\noindent
PACS numbers:  14.60.Pq, 95.30.Cq, 97.10.Cv, 97.60.Bw,
98.65.Dx

\vfil\eject

In this paper we examine constraints on the possible
spectrum of
neutrino masses in light of recent experiments and
astrophysical
models. Surprisingly, and perhaps disturbingly, we find that
if we
insist on having a significant amount of light neutrino dark
matter,
and assume that $r$-process nucleosynthesis originates in
neutrino-heated Type II supernova ejecta, then standard
interpretations of the experiments force us into adopting an
inverted
neutrino mass hierarchy.

Three experiments, or sets of experiments, have been widely
interpreted as suggesting evidence for neutrino oscillations
and,
therefore, the existence of neutrino mass. We discuss each
of these
in turn.

(1) The recent report of evidence for
${\bar{\nu}}_\mu\rightleftharpoons {\bar{\nu}}_e$
oscillations found by
the LSND experiment at Los Alamos is
very exciting [1]. This result by itself, if correct, seems
to be
consistent with a fair range in ${{\delta}m^2_{e\mu}}
\approx |
m^2_{{\nu}_{\mu}} - m^2_{{\nu}_e} |$, which is reported [1]
to be
$0.2\ {\rm eV}^2 {\
\lower-1.2pt\vbox{\hbox{\rlap{$<$}\lower5pt\vbox{\hbox{$\sim
$}}}}\ }
{{\delta}m^2_{e\mu}} {\
\lower-1.2pt\vbox{\hbox{\rlap{$<$}\lower5pt\vbox{\hbox{$\sim
$}}}}\ }
20\ {\rm eV}^2 $ (the notation $\delta m^2_{e\mu}$ and
similar notation hereafter are adopted for convenience, and
should be interpreted as the mass-squared differences
between the appropriate neutrino mass eigenstates).
Apparently, the LSND experiment can also detect ${\nu}_\mu
\rightleftharpoons {\nu}_e$ oscillations. When this data is
combined with the antineutrino oscillation signal, it is
reported
that the \lq\lq best fit\rq\rq\ to mass-squared difference
and vacuum
mixing angle could fall in a range around
${\delta}m^2_{e\mu}
\approx 6\ {\rm eV}^2$ and $\sin^2 2{{\theta}_{e\mu}}
\approx 6
\times {10}^{-3}$.  However, it is clear that the
results reported from the LSND experiment leave us quite far
from any real certainty
in the range of ${\delta}m^2_{e\mu}$ and $\sin^2
2{{\theta}_{e\mu}}$.

(2) The apparent deficit in the expected flux of solar
neutrinos has
been confirmed by a number of experiments [2]. Whether a
solution to
this problem demands new neutrino physics, or an alteration
of the solar model, has been extensively
debated [3]. However, with the new calibration of the GALLEX
experiment, and the identification of the principal neutrino
deficit
as originating near the $^7$Be($e^-,\nu_e)^7$Li electron
capture line,
a consensus is building that
neutrino oscillations are at the root of the solution [4].
Either
matter-enhanced Mikheyev-Smirnov-Wolfenstein (MSW) neutrino
flavor
transformation [5], or large mixing angle vacuum neutrino
oscillations [6], involving the electron neutrino could
explain the solar
neutrino deficit. Perhaps the best choices of neutrino
mixing
parameters for solving this problem are those for the
so-called
\lq\lq small angle solution,\rq\rq\ with
${\delta}m^2_{e\alpha}
\approx {10}^{-6}\ {\rm eV}^2$ to ${10}^{-5}\ {\rm eV}^2$
and $\sin^2
2{{\theta}_{e\alpha}} \sim 5\times{10}^{-3}$. Here the
subscript
$\alpha$ refers to a neutrino species ${\nu}_{\alpha}$,
which for the
small angle solution could be either ${\nu}_{\mu}$,
${\nu}_{\tau}$,
or a sterile species ${\nu}_s$. The \lq\lq large mixing
angle
solution\rq\rq  and vacuum oscillation solution encompass
other ranges
of vacuum neutrino properties, ${\delta}m^2_{e\alpha}
\approx
{10}^{-6}\ {\rm eV}^2$ to ${10}^{-4}\ {\rm eV}^2$and $\sin^2
2{{\theta}_{e\alpha}} {\
\lower-1.2pt\vbox{\hbox{\rlap{$>$}\lower5pt\vbox{\hbox{$\sim
$}}}}\ }
0.4$, or ${\delta}m^2_{e\alpha} \sim {10}^{-10}\ {\rm eV}^2$
and
$\sin^2 2{{\theta}_{e\alpha}} {\
\lower-1.2pt\vbox{\hbox{\rlap{$>$}\lower5pt\vbox{\hbox{$\sim
$}}}}\
}0.75$, respectively. In neither one of these latter two
solutions
could ${\nu}_{\alpha}$ be a sterile species [6,7].

(3) High energy cosmic rays incident on the upper atmosphere
produce
large numbers of $\pi^+$ and $\pi^-$. The decay of these
particles
produces $\nu_e$, $\bar\nu_e$, $\nu_\mu$, and $\bar\nu_\mu$,
which
should occur in the ratio $(\nu_\mu + \bar\nu_\mu)/(\nu_e +
\bar\nu_e) \approx 2$. Instead, Kamiokande, and perhaps
other
experiments, observe this ratio to be close to unity [8].
This
apparent deficit itself, together with the zenith angle
dependence of
the ratio of muon-type to electron-type neutrinos, has been
argued [8,9] to be evidence for $\nu_\mu \rightleftharpoons
\nu_\beta$
vacuum oscillations with $\delta m^2_{\mu\beta} \sim
{10}^{-3}\ {\rm eV}^2$ to
${10}^{-1}\ {\rm eV}^2$. Here $\nu_\beta$ probably could be
only
$\nu_e$ or $\nu_\tau$, as the vacuum mixing angle required
to explain
the data is very large, $\sin^2 2\theta_{\mu\beta} {\
\lower-1.2pt\vbox{\hbox{\rlap{$>$}\lower5pt\vbox{\hbox{$\sim
$}}}}\ }
0.4$. A sterile candidate for $\nu_\beta$
having such a large mixing with $\nu_\mu$
would be disallowed
because it would increase the primordial
$^4$He abundance
by increasing the number of degrees of freedom
extant at the epoch of primordial nucleosynthesis [7].

If all of the above experiments, and their interpretations
in terms
of neutrino oscillations, are correct then it is clear that
$\nu_\alpha$ cannot be the same as $\nu_\beta$. This is
because the
three distinctive mass splittings implied by these
experiments do not
overlap. Furthermore, adoption
of the putative LSND result forces us to conclude that there
is no
mutually consistent identification for $\nu_\alpha$ and
$\nu_\beta$
without the introduction of a sterile neutrino species!
The LSND limit $\delta m^2_{e\mu} > 0.2\ {\rm eV}^2$ implies
that $\nu_\beta
\ne \nu_e$, so $\nu_\beta=\nu_\tau$; it also implies that
$\nu_\alpha \ne \nu_\mu$, and
$\nu_\alpha=\nu_\tau$ is then inconsistent with (3), so
$\nu_\alpha=\nu_s$. This conclusion remains true, even if we
adopt a rigorous treatment for the mixing of three or more
neutrinos.

Two additional astrophysical arguments may provide yet a
different
set of insights into the possible masses and mixings of
light
neutrinos. These astrophysical considerations are: (a) the
possible
requirement of a significant contribution to the closure
density of the
universe from
light neutrino dark matter; and (b) the best proposed site
for the
synthesis of the $r$-process elements is the neutrino-heated
ejecta
from Type II supernovae. We consider each of these arguments
in turn.

Recent simulations of the evolution of structure in the
early
universe, together with the observations of anisotropy in
the cosmic
microwave background, and observations of the distribution
of
galaxies and hydrogen clouds at high red shift, have been
interpreted
as suggesting the need for a mixture of cold dark matter and
at least
some hot dark matter [10,11]. Light neutrino dark matter
could
certainly suffice for the suggested hot dark matter
component. In
fact, the fraction of the closure density, $\Omega_\nu$,
contributed
by the sum of the light neutrino masses,
$\sum_{i}{m_{\nu_i}}$, would
be
$$
\Omega_\nu \approx 0.053 \left(
{{\sum_{i}{m_{\nu_i}}}\over{5\ {\rm
eV}}}\right) {h^{-2}},\eqno(1)
$$
where $h$ is the Hubble parameter in units of $100\ {\rm
km}\ {\rm
s}^{-1}\ {\rm Mpc}^{-1}$, and where the sum on neutrino
masses runs
over three flavors ($\nu_e$, $\nu_\mu$, $\nu_\tau$) and does
not
include sterile species. Recent attempts to reconcile the
results of
large scale structure evolution computations with the
observational
data have led researchers to suggest that
$\sum_{i}{m_{\nu_i}}
\approx 5\ {\rm eV}$ is preferred [11]. Although the
specific scheme of
Primack et al. [11] utilizes $m_{\nu_\tau} \approx
m_{\nu_\mu}
\approx 2.4\ {\rm eV}$, we note that the cosmological
aspects of
their results do not depend on which neutrino flavors have
the
requisite mass, or even how the mass is divided between
them. However,
sharing the mass between two or three neutrino flavors
improves the fit to the
observed number density of galaxy clusters.

The problem of understanding galaxy formation and the large
scale
distribution of matter in the universe is vexing and
complicated.
Nevertheless, mixed cold plus hot dark matter models make
unique and
ultimately testable predictions on the
evolution of the numbers of low mass systems ({\it e.g.},
galaxies, quasars, damped Ly $\alpha$ clouds) with red
shift. Future
observations may give definitive confirmation or rejection
of these
models, but we feel that they provide at least a viable fit
to the
observations at the present time.

By far the best proposed site for the synthesis of the
neutron-rich
heavy elements ({\it e.g.}, uranium) in the $r$-process
(rapid neutron
capture process) is the neutrino-heated ejecta from the
post-core-bounce environment of Type II supernovae (\lq\lq
hot bubble\rq\rq
) [12]. This putative $r$-process site has the advantage
that it
yields the observed solar system abundance distribution of
$r$-process nuclei, and each
supernova makes an amount of $r$-process material which is
in accord
with models of galactic chemical evolution. No other
proposed
$r$-process site can accomplish these feats without the
introduction of
{\it ad hoc} parameters. Additionally, the conditions which
determine
nucleosynthesis in the post-core-bounce \lq\lq hot
bubble\rq\rq\
environment are expected to be independent of the messy
details of
the supernova explosion mechanism.  The conditions conducive
to the
$r$-process will arise in the late stages of all successful
Type II
supernovae which leave  hot neutron star remnants.

However, it has been shown that matter-enhanced neutrino
flavor
transformation ($\nu_e \rightleftharpoons \nu_{\mu(\tau)}$
or
$\bar\nu_e \rightleftharpoons \bar\nu_{\mu(\tau)}$) can
affect
supernova dynamics and nucleosynthesis [13]. In fact,
$r$-process
nucleosynthesis from neutrino-heated supernova ejecta cannot
occur
unless the material in the hot bubble has an excess of
neutrons over
protons. In turn, the neutron-to-proton ratio in this
environment is determined by the spectra of the $\bar\nu_e$
and
$\nu_e$. These facts have been used to place broad limits on
the
mixing parameters of a light $\nu_e$ with $\nu_\mu$ and/or
$\nu_\tau$ possessing
cosmologically significant masses [14].

Detailed numerical calculations which include the
neutrino-neutrino
scattering contributions to the neutrino effective masses
confirm that $r$-process nucleosynthesis is sensitive to
neutrino
flavor mixing [15,16]. In fact, the studies in Refs. [14],
[15], and [16] show
that the \lq\lq best fit\rq\rq\ LSND parameters,
${\delta}m^2_{e\mu}
\approx 6\ {\rm eV}^2$ and $\sin^2 2{{\theta}_{e\mu}}
\approx 6
\times {10}^{-3}$, are not consistent with the $r$-process
originating
in neutrino-heated supernova ejecta if the vacuum neutrino
masses
satisfy $m_{\nu_\mu} > m_{\nu_e}$. Given the above LSND
parameters, this neutrino mass hierarchy guarantees
matter-enhanced
$\nu_e \rightleftharpoons \nu_\mu$ transformation in the hot
bubble,
with resulting hardening of the $\nu_e$ spectrum, and
consequent
reduction of the neutron fraction below acceptable levels.
References
[14], [15], and [16] show that hot bubble $r$-process
nucleosynthesis could only
be compatible with the reported LSND results if $\delta
m^2_{e\mu} <
2\ {\rm eV}^2$.

However, it has been shown that matter-enhanced $\bar\nu_e
\rightleftharpoons \bar\nu_\mu$ transformation in the hot
bubble will
not result in a proton excess. In this case, there are no
obvious
conflicts with $r$-process
nucleosynthesis [17]. A necessary condition for
matter-enhanced
antineutrino flavor transformation is that there be an
inverted
hierarchy of neutrino masses, $m_{\nu_e} > m_{\nu_\mu}$. If
this
inverted mass scheme obtains, then $r$-process
nucleosynthesis would be
compatible with all LSND parameters.

It is not completely clear, however, whether such
matter-enhanced
antineutrino transformation with the LSND parameters would
yield a $\bar\nu_e$ spectrum compatible with that inferred
from the
SN1987A data taken by the IMB and Kamiokande detectors [18].
These
detectors were known to be sensitive to primarily
$\bar\nu_e$ through
the reaction $\bar\nu_e + {\rm p} \rightarrow {\rm n} +
e^+$.
Antineutrino transformation would certainly increase the
average energy
of $\bar\nu_e$ over
the standard case with no flavor transformation. However,
the
inferred range of temperature for the $\bar\nu_e$ energy
distribution in
Ref. [18] is marginally compatible with $\bar\nu_e
\rightleftharpoons
\bar\nu_\mu$, within the statistical uncertainties of the
SN1987A
data. Furthermore, the underlying neutrino emission model on
which
the analysis of Ref. [18] and similar studies are based, is
now
known to be incorrect ({\it e.g.}, they assume that the
radius of the
neutron star is fixed, the neutrino spectra are black body,
these
spectra cool with time, and the luminosity of each neutrino
species
is given by the first three assumptions --- all of these
points
are wrong). Therefore, any constraints on antineutrino
transformation derived
from SN1987A would be, at best, model dependent. Recent
work by Mayle and Wilson [19] which incorporates
matter-enhanced
antineutrino transformation with the LSND \lq\lq best
fit\rq\rq\
parameters yields a $\bar\nu_e$ spectrum which would give an
acceptable signal in the IMB or Kamiokande detectors for
SN1987A. However,
future water \v{C}erenkov neutrino detectors ({\it e.g.,}
Super
Kamiokande) would certainly see a clear signal for
matter-enhanced
antineutrino oscillations for a galactic supernova.

How seriously should we take these $r$-process
considerations? It
should be borne in mind that the calculations in Refs. [14],
[15], and [16] have
some limitations and caveats. First, we do not know with
complete
certainty where the $r$-process originates. Second, studies
of neutrino
flavor transformation in the post-core-bounce supernova
environment
have only examined two-neutrino mixing [13--16]. If, for
example, the $\nu_\tau$ and $\nu_\mu$, which have identical
energy
spectra in the supernova, also had nearly degenerate masses
and mixed
with the $\nu_e$ in a similar fashion, then for a given
neutrino
energy and $\delta m^2_{e\mu{(\tau)}}$, we would expect the
resonance
regions for $\nu_e \rightleftharpoons \nu_\mu$ and $\nu_e
\rightleftharpoons \nu_\tau$ to be overlapping. Is it then
possible
to get destructive quantum interference of the neutrino
flavor
conversion amplitudes at resonance? These amplitudes are
known to be
energy dependent [15], so that destructive interference in
some
narrow energy region of the neutrino spectrum would be
countered with
constructive interference in another. In this way it is
clear that
one could not engineer destructive interference-induced
reduction in
the degree of neutrino flavor transformation over the whole
neutrino
energy spectrum. Nevertheless, this issue bears further
examination.

If we adopt as correct experiments (1), (2), and (3), along
with
their interpretations in terms of neutrino oscillations, and
adopt
the astrophysical arguments (a) and (b), then there is a
fairly
unique set of neutrino masses and mixings which emerges as a
fit. In
this fit, the masses of the neutrinos must satisfy
$m_{{\nu}_e}
\approx m_{{\nu}_s} \approx 2.7\ {\rm eV}$ (where ${\nu}_e$
is split
from a sterile species, ${\nu}_s$, by $\sim {10}^{-5} \ {\rm
eV}$)
and $m_{{\nu}_{\tau}} \approx m_{{\nu}_{\mu}} \approx 1.1\
{\rm eV}$
(where these species are split by $\sim {10}^{-2} \ {\rm
eV}$). Here, and in what follows, when we say one neutrino
species is split from the other by a certain amount, we mean
that the corresponding neutrino vacuum mass eigenvalues
differ by this amount. In
this scheme, we would adopt $\sin^2 2\theta_{e\mu} \sim
{10}^{-2}$,
$\sin^2 2\theta_{\mu\tau} \sim 1$, and $\sin^2 2\theta_{es}
\sim
{10}^{-2}$. Here we have adopted the \lq\lq best fit\rq\rq\
LSND
result and we assume that $\sum_{i}{m_{\nu_i}} \approx
5\,{\rm eV}$.
This mass spectrum could be altered in obvious fashion if we
adopt
slightly different values for the LSND results and
$\sum_{i}{m_{\nu_i}}$. Note that the production of $\nu_s$
in the early
universe with $\sin^22\theta_{es}\sim10^{-2}$ is negligible,
as would
be required from considerations of big bang nucleosynthesis
and the
observed $^4$He abundance [7]. This justifies the exclusion
of $\nu_s$ in the sum $\sum_im_{\nu_i}$.

The absence of an observation of neutrino-less double beta
decay has
been argued to place a limit on the Majorana mass of the
$\nu_e$ of
$\sim 1\ {\rm eV}$ (actually, a weighted sum of the masses
of all light
neutrino species is constrained to be less than $\sim 1$ eV,
but the
$\nu_e$ usually makes the biggest contribution to the sum)
[20]. With
this limit, it is clear that
the neutrino masses in the above scheme would have to be
Dirac, unless
there were a fortuitous cancellation in the weighted sum
over neutrino
masses. From the tritium end-point experiments, the current
upper limit
on the mass (Dirac or Majorana) of the $\nu_e$ is 7.2 eV
[21].

Note that it is the LSND result which forces us to
contemplate an
inverted neutrino mass hierarchy and/or the introduction of
a sterile
neutrino species. If we were to adopt the LSND result, but
give up atmospheric
neutrino oscillations and the hot bubble $r$-process, then
we could
have $m_{\nu_e} \approx m_{\nu_\tau} \approx 1.1\ {\rm eV}$
and
$m_{\nu_\mu} \approx 2.7\ {\rm eV}$, with $\nu_e$ and
$\nu_\tau$
split by $\sim {10}^{-5}\ {\rm eV}$, a value sufficient to
give an
MSW solution for the solar neutrino problem (assuming
$\sin^22\theta_{e\tau}\sim10^{-2}$). This scheme has the
obvious advantage that it does not require the introduction
of a
sterile neutrino species, but we note that it does have
a curious
neutrino mass hierarchy (essentially, this is an inverted
neutrino mass hierarchy, since the second family $\nu_\mu$
is heavier than the third family $\nu_\tau$). This scheme
could be consistent with either
Majorana or Dirac neutrino masses. If we were to modify this
scheme
by requiring atmospheric neutrino oscillations, but dropping
the MSW mechanism in
the sun, then we could have $m_{\nu_\mu} \approx
m_{\nu_\tau} \approx
2.5\ {\rm eV}$ (with these species split by $\sim {10}^{-2}\
{\rm
eV}$ and $\sin^2 2\theta_{\mu\tau} \sim 1$), and $m_{\nu_e}
\sim 0$
with $\sin^2 2\theta_{e\mu} \sim {10}^{-2}$, with no mass
inversion
and no sterile species. We could restore the MSW mechanism
in the sun for this
latter scheme by simply adding a sterile neutrino species
split from
the $\nu_e$ by $\sim {10}^{-3}\ {\rm eV}$ with $\sin^2
2\theta_{es}
\sim {10}^{-2}$. This scheme then accounts for all
constraints, except for
the $r$-process.

It is tempting to visit a scheme where $m_{\nu_e} \approx
m_{\nu_\tau} \approx 2.5\ {\rm eV}$, with these species
split by $\sim
{10}^{-6}\ {\rm eV}$, to give the MSW mechanism in the sun
(assuming
$\sin^22\theta_{e\tau}\sim10^{-2}$), and where $m_{\nu_\mu}
\approx m_{\nu_s} \approx 0$, with these species split by
$\sim
{10}^{-1}\ {\rm eV}$ to $\sim{10}^{-2}\ {\rm eV}$, to give
atmospheric
neutrino oscillations. Note, however, that the large mixing
between
$\nu_\mu$ and $\nu_s$ required for atmospheric neutrino
oscillations,
$\sin^2 2\theta_{\mu{s}} \sim 1$, is probably precluded by
big bang
nucleosynthesis considerations [7]. If we drop the sterile
neutrino
species from this scheme, then we can explain all of the
above
constraints (1, 2, a, b), except for the atmospheric
neutrino deficit.
Since the explanation of the atmospheric neutrino deficit in
terms of neutrino oscillations is far from settled, this may
be an
attractive scenario if the LSND result holds up.

\centerline{\bf ACKNOWLEDGMENTS}

The authors acknowledge discussions with D. Caldwell, W.
Haxton, A. Klypin, R.
Mayle,
W. Vernon, J. R. Wilson, and S. E. Woosley. This work was
supported
by the Department of Energy under Grant No.
DE-FG06-90ER40561 at the
Institute for Nuclear Theory,
by NSF grant PHY-9121623 at UCSD, and by NSF grant
PHY-9402455 at UCSC.

\vfil\eject
\vskip 0.2in\noindent
\centerline{\bf REFERENCES}
\vskip 0.2in
\nopagenumbers
\noindent
[1] LSND, New York Times, January 31, 1995. Also
presentation
by D. Hywel White, TUNL meeting on electroweak physics,
astrophysics,
and nonaccelerator physics, LBL, Berkeley, CA, February 4,
1995.\hfil\break
\noindent
[2] Cf. the review by R. S. Raghavan, Science {\bf
267}, 45 (1995), and references therein.\hfil\break
\noindent
[3] J. N. Bahcall and M. Pinsonneault, Rev. Mod. Phys. {\bf
64}, 885 (1992); see also the {\it Proceedings of the Solar
Modeling
Workshop}, Seattle, WA, March 1994 (Institute for Nuclear
Theory,
Seattle, WA, in press).\hfil\break
\noindent
[4] N. Hata, S. Bludman, and P. Langacker, Phys. Rev. D {\bf
49}, 3622
(1994).\hfil\break
\noindent
[5] L. Wolfenstein, Phys. Rev. D {\bf 17}, 2369 (1978); {\bf
20}, 2364
(1979); S. Mikheyev and A. Yu. Smirnov, Nuovo Cimento Soc.
Ital. Fis.
{\bf 9C}, 17 (1986); H. A. Bethe, Phys. Rev. Lett. {\bf 56},
1305
(1986).
\hfil\break
\noindent
[6] P. I. Krastev and S. T. Petcov, Phys. Rev. Lett. {\bf
72}, 1960
(1994).\hfil\break
\noindent
[7] X. Shi, D. N. Schramm, and B. D. Fields, Phys. Rev. D
{\bf 48}, 2563
(1993).\hfil\break
\noindent
[8] Y. Fukuda et al., Phys. Lett. {\bf B335}, 237 (1994); R.
Becker-Szendy et al., Phys. Rev. Lett. {\bf 69}, 1010
(1992).\hfil\break
\noindent
[9] W. Frati, T. K. Gaisser, A. K. Mann, and T. Stanev,
Phys. Rev. D {\bf 48}, 1140 (1993), and references
therein.\hfil\break
\noindent
[10] S. A. Bonometto and R. Valdarnini, Phys. Lett. {\bf
103A}, 369
(1984); L. Z. Fang, S. X. Li, S. P. Xiang, Astron.
Astrophys. {\bf 140},
77 (1984); A. Dekel and S. J. Aarseth, Astroph. J. {\bf
283}, 1 (1984);
Q. Shafi and F. W. Stecker, Phys. Rev. Lett. {\bf 53}, 1292
(1984).\hfil\break
\noindent
[11] A. Klypin, S. Borgani, J. Holtzman, J. R. Primack,
Astrophys. J.,
in press (1995);
J. R. Primack, J. Holtzman, A. Klypin, D. O. Caldwell, Phys.
Rev.
Lett., in press
(1995); D. Pogosyan and A. Starobinsky, Astrophys. J., in
press  (1995).
\hfil\break
\noindent
[12] S. E. Woosley, J. R. Wilson, G. J. Mathews, R. D.
Hoffman, and B.
S. Meyer, Astrophys. J. {\bf 433}, 229 (1994).\hfil\break
\noindent
[13] G. M. Fuller, R. Mayle, B. S. Meyer, and J. R. Wilson,
Astrophys.
J. {\bf 389}, 517 (1992); G. M. Fuller, Phys. Rep. {\bf
227}, 149 (1993).
\hfil\break
\noindent
[14] Y.-Z. Qian, G. M. Fuller, R. W. Mayle, G. J. Mathews,
J. R.
Wilson, and S. E. Woosley, Phys. Rev. Lett. {\bf 71}, 1965
(1993).\hfil\break
\noindent
[15] Y.-Z. Qian and G. M. Fuller, Phys. Rev. D {\bf 51}, in
press
(1995).\hfil\break
\noindent
[16] G. Sigl, Phys. Rev. D, in press
(1995).\hfil\break
\noindent
[17] Y.-Z. Qian and G. M. Fuller, Phys. Rev. D, submitted
(1995); Y.-Z. Qian, PhD Thesis, University of California,
San Diego
(1993).\hfil\break
\noindent
[18] T. J. Loredo and D. Q. Lamb, in {\it Proceedings of the
14th Texas
Symposium on Relativistic Astrophysics}, edited by E.
Fenyves (New York
Academy of Sciences 1989).\hfil\break
\noindent
[19] R. W. Mayle and J. R. Wilson, private communication
(1995).\hfil\break
\noindent
[20] See, for example, W. C. Haxton and G. J. Stephenson,
Prog. Part.
Nucl. Phys. {\bf 12}, 409 (1984); P. Vogel and M. R.
Zirnbauer, Phys.
Rev. Lett. {\bf 57}, 3148 (1986); and A. Balysh et al.,
Phys. Lett. {\bf
B283}, 32 (1992).  \hfil\break
\noindent
[21] Ch. Weinheimer et al., Phys. Lett. {\bf B300}, 210
(1993).\hfil\break
\vfil\eject
\end